# Strain Correlated Linearly Polarized Photoluminescence in $WS_2/WSe_2$ Moiré Superlattices

*Yuto Urano[1,2], Ryo Tamura[3,4], Yui Tamogami[1,5], Toshikaze Kariyado[1], Yasumitsu Miyata[1], Daichi Kozawa[1], Kenji Watanabe[6], Takashi Taniguchi[1], and Ryo Kitaura[1,2,*]*

[1] *Research Center for Materials Nanoarchitectonics, National Institute for Materials Science, 1-1 Namiki, Tsukuba, Ibaraki 305-0044, Japan*

[2] *Graduate School of Chemical Sciences and Engineering, Hokkaido University, 5, Kita 8 Nishi, Kita-ku, Sapporo, Hokkaido 060-8628, Japan*

[3] *Center for Basic Research on Materials, National Institute for Materials Science, 1-1 Namiki, Tsukuba, Ibaraki 305-0044, Japan*

[4] *Graduate School of Frontier Sciences, The University of Tokyo, 5-1-5 Kashiwanoha, Kashiwa, Chiba 277-8568, Japan*

[5] *Department of Physics, Tokyo Metropolitan University, 1-1 Minamiosawa, Hachioji, Tokyo 192-0397, Japan*

[6] *Research Center for Electronic and Optical Materials, National Institute for Materials Science, 1-1 Namiki, Tsukuba, Ibaraki 305-0044, Japan*

* Correspondence to KITAURA.Ryo@nims.go.jp

**ABSTRACT**

Reliable optical control of valley degrees of freedom in moiré excitons requires that the emitted polarization faithfully reflect the underlying valley state. Here, we show that linearly polarized photoluminescence from $WSe_2/WS_2$ moiré excitons is largely insensitive to the excitation linear polarization and therefore does not arise from valley coherence. Automated polarization-resolved photoluminescence and Raman mapping at cryogenic temperature reveal that the degree of linear polarization correlates strongly with local Raman shifts and moiré-exciton observables, identifying strain as the dominant experimental correlate. Linear-regression analysis further shows that strain-related descriptors provide the best prediction of the observed polarization. Guided by theory, we attribute this behavior to strain-amplified breaking of $C_3$ symmetry in the moiré potential: weak uniaxial strain produces only partial cancellation of locally elliptical emission, yielding a finite far-field degree of linear polarization. These results establish strain as a key control parameter for reliable optical readout in TMD moiré superlattices.

*Keywords: moiré superlattices; transition metal dichalcogenides; excitons; strain; polarization-resolved photoluminescence; valleytronics*

## INTRODUCTION

The interplay between valley degrees of freedom (VDF) and light polarization represents a critical aspect of transition metal dichalcogenides (TMDs)[1–4]. Semiconducting TMDs, such as $WS_2$ and $WSe_2$ in their 2H configuration, exhibit a bandgap at the K and −K points located at the corners of the Brillouin zone[5,6]. In monolayer TMDs, the breaking of inversion symmetry results in the K and −K points being non-equivalent, and thus the valley index must be specified to describe carriers and excitons in these systems. Because of the optical selection rule arising from three-fold rotational symmetry, excitons at K and −K selectively couple with right and left circularly polarized light, leading to optical generation and detection of VDF of excitons.[7,8] The VDF-polarization coupling in TMDs allows us to optically control VDF for next-generation valleytronic devices[9,10].

The VDF-polarization coupling is also important in moiré superlattices of TMDs. In the moiré superlattices, such as $WSe_2/WS_2$ and twisted $MoSe_2/WSe_2$, interlayer moiré excitons emerge due to the type-II band alignment[11–13]. Because the electron and hole are spatially separated, interlayer moiré excitons exhibit a long lifetime.[14–16] Additionally, due to the local structures with threefold rotational symmetry ($C_3$ symmetry) in moiré superlattices, moiré exciton wave packets confined in these high-symmetry locations, A, B, and C, follow $C_3$ symmetry at K and −K points, leading to the valley-selective optical selection rule coupled with circularly polarized light [17–20]. The long lifetimes and optical selection rules of moiré excitons enable control of VDF via optical generation and detection.

The potential for optical control relies on a one-to-one correspondence between VDF and light polarization. Circularly polarized light corresponds to the K and −K valleys, while linearly polarized light, a superposition of left and right circularly polarized light, corresponds to a superposition of the K and −K valleys[21–25]. This one-to-one correspondence enables the preparation of desired valley states and their optical readout. When combined with optical controls such as the valley-selective optical Stark effect, the VDF-polarization correspondence could enable quantum technologies that rely on all-optical generation, manipulation, and detection.

In this work, however, we demonstrate that, unlike monolayer $WSe_2$, where the linear polarization in photoluminescence (PL) follows the excitation linear polarization, $WSe_2/WS_2$ moiré exciton shows linearly-polarized PL insensitive to the excitation linear polarization, indicating it does not originate from coherent superposition of K and −K valleys. Polarization-resolved PL and Raman maps reveal that the linear polarization correlates with strain across the sample. Guided by a theoretical calculation, we suggest that even a weak strain can distort the moiré potential and weakly break $C_3$ symmetry, yielding a non-vanishing linear (elliptical) component in the far-field polarization. These findings reveal the strain sensitivity of moiré excitons and highlight an essential consideration for valleytronic device designs utilizing TMD moiré superlattices.

## RESULTS AND DISCUSSION

Figure 1a shows an optical image of a triangular $WS_2$ crystal, and smaller triangular $WSe_2$ crystals grown on the $WS_2$ crystal are visible in the inset PL image. In this combination, due to the differences in lattice parameters of each layer, a moiré superlattice can form even when the twist angle between $WS_2$ and $WSe_2$ is zero. One benefit of direct growth is the clean interfaces between $WS_2$ and $WSe_2$ with a well-defined interlayer twist angle of 0 deg. Figures 1b and 1c present room temperature PL and Raman spectra of a $WS_2/WSe_2$ heterostructure grown on an hBN flake. The Raman spectrum measured with 632.8 nm excitation exhibits peaks at 420, 350, and 250 $cm^{-1}$, which can be assigned to $A'_1$, $E^{'}$ modes of $WS_2$ and $A'_1$ modes of $WSe_2$[25–28]. A room temperature PL spectrum shows peaks at 2.12 and 1.67 eV, which arise from 1s exciton resonance of $WS_2$ and $WSe_2$, respectively[29–32]. A PL spectrum measured at 3.5 K (Fig. 1d) shows low-energy peaks that are absent in the room temperature PL, and these peaks are characteristic of moiré excitons in this system[33-34]. All Raman and PL spectroscopy results discussed above are well consistent with the formation of a $WSe_2/WS_2$ moiré superlattice.

Accurate excitation with polarized light and PL polarization measurements are crucial for controlling VDF in TMD-based heterostructures. To achieve this, we utilized an optical system, as illustrated in Fig. 2a. The polarizer and half-wave plate, located directly below the beam splitter, determine the polarization of the excitation light — either vertical or horizontal — before the beam is focused onto the sample. The PL emitted from the sample passes directly to the other half-wave plate and polarizer without any mirrors or optical devices that could disrupt its polarization. Additionally, we incorporated a He-Ne laser for measuring Raman spectra in addition to PL spectra. The control of the sample stage, rotation of half-wave plates, and the on/off operation of the shutters for both the Ti-sapphire and He-Ne lasers are automated in our optical system, enabling fully automated polarization-resolved PL and Raman mapping of samples.

Before measuring the $WSe_2/WS_2$ moiré sample, we conducted polarization-resolved PL spectroscopy of monolayer $WSe_2$. As shown in the left-hand side of Fig. 2b, the PL from $WSe_2$ is linearly polarized, exhibiting a degree of linear polarization (DLP) of 18.8 %. The DLP is defined as $(I_0 - I_{90}) / (I_0 + I_{90})$, where $I_0$ and $I_{90}$ represent the PL intensities with polarizations parallel and perpendicular to that of the excitation light, respectively. To determine whether the origin of the DLP stems from a superposition of K and −K valleys, we also measured the light polarization of PL using horizontal excitation. As shown in the figure, the polarization direction rotates by 90 degrees when it is switched from vertical to horizontal. The change in polarization direction clearly shows that the emission polarization aligns with the excitation polarization, consistent with the generation and detection of the K and −K superposition states using linearly polarized light. The DLP of 18.8 % is also consistent with previously reported values[35,36].

In contrast, the polarization of PL from the $WSe_2/WS_2$ moiré sample is insensitive to the excitation light polarization. As illustrated on the right-hand side of Fig. 2b, the polarization direction remains unchanged regardless of the excitation polarization; it does not align with either the vertical or the horizontal direction. The DLP map shown in Fig. 2c demonstrates that a finite DLP of 3 % to 7 % is present over a broad range of the sample, while the rotation of the polarization direction by altering the excitation polarization is nearly zero in all places (Fig. 2d). The polarized PL that is insensitive to excitation polarization differs significantly from those observed in $WSe_2$, indicating it likely originates from a different source other than VDF.

To address the origin of light polarization in PL, we have conducted polarization-resolved PL and Raman mapping of the sample. It is important to note that mapping measurements can take one to two days, and drift correction is necessary for obtaining accurate spectral mapping (see Fig. S1 for details). In addition to the maps shown in Figs. 2c and d, we produced maps for six quantities, including Raman $A'_1$ mode of $WS_2$ and $WSe_2$ ($A'_1$ ($WS_2$) and $A'_1$ ($WSe_2$)), moiré exciton PL intensity ($I$(moiré)), moiré exciton PL peak position ($E$(moiré)), $WSe_2$ intralayer trion intensities ($I(X^-)$), polarization direction of moiré exciton PL ($\theta$(axis)). As shown in Figs. 3a-h, although the $WSe_2/WS_2$ moiré sample was prepared through direct CVD growth, variations in values across all maps indicate inhomogeneous strain and doping. The Raman maps (Fig. 3a and b) are somewhat similar to the $I(X^-)$ map (Fig. 3e), likely due to frequency shifts of Raman modes induced by carrier doping. As illustrated in Fig. 3a and b, the DLP map and Raman maps, particularly the $WS_2$ $A'_1$ map, exhibit significant similarities, suggesting that the DLP observed in the $WSe_2/WS_2$ moiré sample arises from strain within the sample.

To quantitatively analyze map similarities, we calculated Pearson correlation coefficients, which measure the linear correlation between two sets of properties. Figure 3g displays pairwise Pearson correlation coefficients among eight quantities, including $I(X^-)$, $A'_1$ ($WS_2$), $A'_1$ ($WSe_2$), $\theta$ (axis), $E$(moiré), $E(WSe_2)$, $I$(moiré), and DLP. Note that $E$(moiré) is defined by the average positions of three peaks observed in the moiré exciton region. All diagonal elements are equal to one because the linear correlation of identical properties should equal one by the definition of the Pearson correlation coefficient. Both Raman modes of $WSe_2$ and $WS_2$ should be affected by local strain and doping, and the correlation coefficient of 0.76 between them is reasonably high. Another observation is the positive correlation between the $I(X^-)$ and Raman shifts, as evidenced by the similarity between the $I(X^-)$ and Raman shift maps. The $I(X^-)$ represents the carrier density in the sample, and the positive correlation coefficients indicate that an increase in carrier density causes a blue shift in the $A'_1$ and $A'_1$ Raman frequencies[37,38].

The DLP exhibits correlation coefficients of −0.64 and −0.44 with the Raman modes, reflecting the similarities between the DLP map and the Raman mode maps. The correlation with Raman modes suggests that the observed DLP is associated with strain in the sample. Additionally, the correlation coefficient with $E$(moiré) is also high, 0.70. This aligns with the strain-induced DLP, as the strain is also expected to influence the moiré exciton peak positions through strain-induced moiré potential modulation (a related discussion is provided below). Also, as discussed later, $\theta$(axis) is directly linked to the deformation of the strain-induced moiré potential, resulting in a high correlation of −0.75. In contrast, the correlation coefficients with $I(X^-)$ and the peak positions of the intralayer exciton are pretty low. The low coefficient for $I(X^-)$ indicates that the dependence of the DLP does not arise from local Fermi level fluctuations in the $WSe_2/WS_2$ sample. The fluctuation of the Fermi level can locally modulate the carrier density, altering the dynamics of excitons, including trion formation and Auger recombination. The observed low correlation indicates that the PL polarization is not directly related to these dynamics.

Based on the experimental results, we have constructed linear regression models for DLP prediction. To perform the feature selection, the prediction accuracy was calculated for all combinations of properties using 10-fold cross-validation[39] (we have investigated $2^8$-1=255 combinations, see Fig. S2). The highest prediction accuracy was obtained when using the four quantities: $I(X^-)$, $A'_1$ ($WS_2$), $\theta$(axis), and $E$(moiré). These four quantities are also used in all top-10 prediction models, as illustrated in Fig. S2, underscoring their importance for understanding the observed DLP. Figure 3h represents the relationship between the actual and predicted values for the test data, along with the $R^2$ score. Although the prediction accuracy is not remarkably high, DLP can be predicted using four quantities. Although $A'_1$ ($WS_2$), $\theta$(axis), and $E$(moiré) should be closely correlated with strain, $I(X^-)$, which has a small correlation coefficient, is still included in the prediction model. The weak direct correlation between $I(X^-)$ and DLP indicates that the observed DLP is unlikely to be governed primarily by local carrier-density variations. Nevertheless, $I(X^-)$ is included as a covariate in the regression model because Raman shifts can contain both strain- and carrier-density-related contributions. Thus, $I(X^-)$ is used to control for carrier-density effects rather than to identify the primary origin of DLP.

The $WSe_2/WS_2$ sample used was prepared by direct growth onto an hBN flake. As shown in Fig. S3, the AFM height image reveals a flat surface with small surface roughness, as expected for the sample on hBN flakes with a clean interface. Bubble-like contrasts are also evident in the image, likely due to impurities trapped between the top hBN and the $WSe_2/WS_2$ heterostructure; we placed an hBN overlayer after growth to protect the heterostructure from adsorbed impurities. The impurities, along with the sub-micrometer-scale roughness in the bottom hBN layer—likely resulting from the heating and cooling processes during chemical vapor deposition (CVD) and the differences in thermal expansion coefficients between hBN and $SiO_2$—probably contribute to the strain distribution observed in the Raman shift patterns displayed in Figs 3a and 3b.

In moiré superlattices, slight strain can cause significant distortion of the moiré period, $d_{moiré}$. The $d_{moiré}$ is represented by $d_{WS2}d_{WSe2}/(d_{WSe2}-d_{WS2})$, where $d_{WSe2}$ and $d_{WS2}$ correspond to the cell parameters of $WSe_2$ and $WS_2$, respectively. The $d_{WSe2}$ of 0.328 and the $d_{WS2}$ of 0.315 nm yield[40,41] a $d_{moiré}$ of 7.95 nm, which matches the reported experimental values well. The derivative of $d_{moiré}$ with respect to $d_{WS2}$ is proportional to $(d_{WSe2}/(d_{WSe2}-d_{WS2}))^2$, and the denominator, $(d_{WSe2}-d_{WS2})^2$, is small due to the slight difference between $d_{WSe2}$ and $d_{WS2}$, indicating that a tiny change in $d_{WS2}$ can cause a significant alteration in $d_{moiré}$. For instance, a tensile strain of 0.1% in $WS_2$ results in a 2.6% change in $d_{moiré}$ along the direction of the strain, making it 23 times as large as the original strain. When uniaxial strain is introduced, $d_{moiré}$ can be distorted. Figure 4a displays structural models of $WSe_2/WS_2$ under two conditions: without strain and with 2 % tensile uniaxial strain applied to the $WS_2$ layer along the armchair direction; the Poisson ratio of $WS_2$ is used to construct the structural model[42]. As illustrated, a tensile strain of 2 % in the $WS_2$ layer distorts the moiré potential, transforming its shape from circular to ellipsoidal, with an aspect ratio of 2.39. Figure 4b shows the relationship between uniaxial strain and aspect ratio, highlighting that even minor strains can cause a noticeable distortion of the moiré potential.

The wavefunctions of the conduction and valence bands can be expressed as $exp(i\vec{k}\cdot\vec{r})\chi(r)u(r)$, where $exp(i\vec{k}\cdot\vec{r})$ represents the Bloch phase factor and $\chi(r)$ represents the envelope function, which represents the confinement arising from the moiré potential, and $u(r)$ denotes the Bloch function, mainly composed of d-orbitals of tungsten atoms. While the Bloch phase term just gives momentum conservation in optical transitions, $\chi(r)$ and $u(r)$ can cause polarization selectivity. As previously discussed, even a small strain can significantly distort the moiré potential, causing the shape of $\chi(r)$ to change from symmetric to ellipsoidal. While $\chi(r)$ does not directly give polarization selectivity in the interband transitions, $u(r)$ can contribute polarization selectivity through the transition matrix element $\langle u_c(r)|\hat{p}(\theta)|u_v(r)\rangle$, where $u_c(r)$ and $u_v(r)$ correspond to the Bloch function of the conduction and valence bands, respectively. $\hat{p}(\theta)$ is defined as $\boldsymbol{e}(\theta)\cdot\hat{p}$, where $\boldsymbol{e}(\theta)$ and $\hat{p}$ represent a unit vector, whose direction is determined by $\theta$, and the momentum operator, $-i\hbar\left(\frac{\partial}{\partial x},\frac{\partial}{\partial y}\right)$, respectively. Figure 5 displays a 2D map of the local matrix element along with a corresponding structural model. The $\theta$ dependence, which determines polarization selectivity in optical transitions, is shown as polar plots at each location; the corresponding interlayer configurations are shown on the left. Interestingly, the $\theta$ dependence exhibits a p-like shape, except at the A, B, and C sites, indicating linearly polarized light emission at each location. However, due to the $C_3$ symmetry of the system, the overall polarizations cancel each other out, resulting in PL without a specific linear polarization. When strain induces the $C_3$ symmetry breaking, as evidenced by the moiré pattern distortion, the cancellation of light polarization is incomplete, resulting in a finite residual linear polarization. Moreover, the distortion of $\chi(r)$ induced by strain can further modulate the spread of the exciton wavefunction, contributing to the observed residual linear polarization by affecting the emission-contributing region. Figure 5 should be regarded as a local transition-matrix analysis for finite DLP, not as an evaluation of the circular valley selection rule; see Supporting Note S5 for more explanation.

To clarify the mechanism for the emergence of finite linear polarization without relying on large-scale calculations, we provide a minimal symmetry-based model in Supporting Note S6. In this model, three $C_3$-related local emission channels cancel in the symmetric limit. Strain-induced distortion of the moiré pattern and the exciton envelope is then represented by an effective $C_3$-breaking parameter, which makes the three local emission weights inequivalent and yields a finite far-field DLP that is linear in this parameter to first order. This residual polarization can arise even from weak strain because the strain effect is amplified in moiré superlattices, as discussed above. This finding highlights the need to account for residual strain when interpreting linearly polarized PL from moiré excitons as a readout of valley coherence.

We note that this conclusion does not imply that weak $C_3$ symmetry breaking immediately destroys circularly polarized valley readout. Circular polarization primarily probes the population imbalance between the K and −K valleys, whereas linear polarization probes their coherent superposition. A weak $C_3$-breaking perturbation can generate an additional linear-polarization component through incomplete cancellation of local emission channels, even when the circularly polarized contrast associated with valley polarization remains approximately valid. A simple discussion of this distinction is provided in Supporting Note S7. The present result should thus be regarded as a caution specifically for interpreting finite DLP as evidence of valley coherence in moiré excitons.

## CONCLUSION

In conclusion, our polarization-resolved PL and Raman mapping of $WSe_2/WS_2$ moiré superlattices shows that finite PL DLP from moiré excitons should not be interpreted simply as a signature of valley coherence. Unlike monolayer $WSe_2$, where the PL linear-polarization axis follows the excitation linear polarization through valley-coherent optical selection rules, the moiré heterostructure exhibits a finite DLP whose polarization axis is largely insensitive to the excitation polarization. Spatially resolved PL and Raman mapping, combined with correlation analysis, reveal that the DLP strongly correlates with Raman shifts and moiré-exciton observables, suggesting that unintentional strain is the dominant experimental correlate. We attribute this behavior to strain-amplified distortion of the moiré atomic registry and moiré potential landscape, which modifies the relative weights of local emission channels and the spatial distribution of the exciton envelope. These results show that even weak residual strain can generate valley-coherence-independent linear polarization in moiré exciton emission. Thus, residual strain and moiré-amplified symmetry breaking must be considered when using linearly polarized PL as an optical readout of valley coherence in TMD moiré heterostructures.

## METHODS

### Preparation of $WSe_2/WS_2$ moiré heterostructures

hBN flakes were mechanically exfoliated onto $SiO_2$/Si substrates with a 285-nm-thick $SiO_2$ layer. Monolayer $WS_2$ was first grown on the hBN flakes by powder CVD using $WO_3$ powder, KBr powder as a growth promoter, and sulfur flakes as precursors. The growth was performed under $N_2$ flow at 600 sccm. The substrate and $WO_3$ source were heated to 970 °C over 40-50 min, while the sulfur source was heated to 180 °C for 10 min when the substrate temperature reached 950 °C. After growth, the system was rapidly cooled with an electric fan. $WSe_2$ was subsequently grown on the as-grown $WS_2$/hBN sample by MOCVD using $(t\text{-BuN=})_2W(NMe_2)_2$ and $(t\text{-}C_2H_5)_2Se_2$ as W and Se precursors, respectively. The sample was first heated to 800 °C in an $N_2/H_2$ flow of 679/21 sccm and annealed for 10 min to remove surface impurities. After cooling to 100 °C, the sample was reheated to 700 °C, and the precursors were supplied for 10 min by $N_2$ bubbling under atmospheric pressure. The precursor reservoirs were kept at room temperature during growth. After growth, the system was rapidly cooled with an electric fan. Finally, the directly grown $WSe_2/WS_2$ heterostructure was encapsulated with a top hBN flake.

### Optical measurements

Optical responses were measured using a home-built microspectroscopy system. PL measurements were performed using a Ti:sapphire pulsed laser (Mai Tai HP, Spectra-Physics) operating at 740 nm, with a repetition rate of 80 MHz and a pulse width of 100 fs. The laser beam was focused to a spot approximately 1 μm in diameter. A typical excitation power of ~10μW was used. Raman spectroscopy experiments were performed using a He-Ne continuous-wave laser (HNL050L, Thorlabs) at 632.8 nm, with a focused spot size of approximately 1 μm in diameter, similar to the pulsed laser setup. For both Raman and PL observations, we used a liquid-nitrogen-cooled CCD detector (PyLoN 400, Teledyne Princeton Instruments) and a spectrometer (IsoPlane 320, Teledyne Princeton Instruments). The sample was mounted in an ultra-low vibration cryogen-free optical cryostat (CryoAdvance 50, Montana Instruments) and maintained at a temperature of 3 K. Linear polarization was generated and detected using polarizers (GTH10M-A for excitation and LPNIR100-MP2 for detection, Thorlabs) as well as λ/2 plates (AHWP05M-580 for excitation and AHWP05M-980 for detection, Thorlabs) located on both excitation and detection sides. These optical elements were controlled by automated rotation stages, and all measurements were conducted fully automatically through a homemade LabVIEW program.

**Acknowledgments**

R.K. was supported by JSPS KAKENHI Grant No. JP23H05469, JP22H05458, JP21K18930, and JP20H05664, and JST SCICORP Grant No. JPMJSC2110 and PRESTO Grant No. JPMJPR20A2. K.W. and T.T. acknowledge support from the JSPS KAKENHI Grant Numbers 20H00354, 21H05233, and 23H02052 and World Premier International Research Center Initiative (WPI), MEXT, Japan.

**Conflicts of Interest**

The authors declare that there is no conflict of interest regarding the publication of this article.

**Data availability**

The data supporting the plots in this paper and the findings of this study are available from the corresponding author upon request.

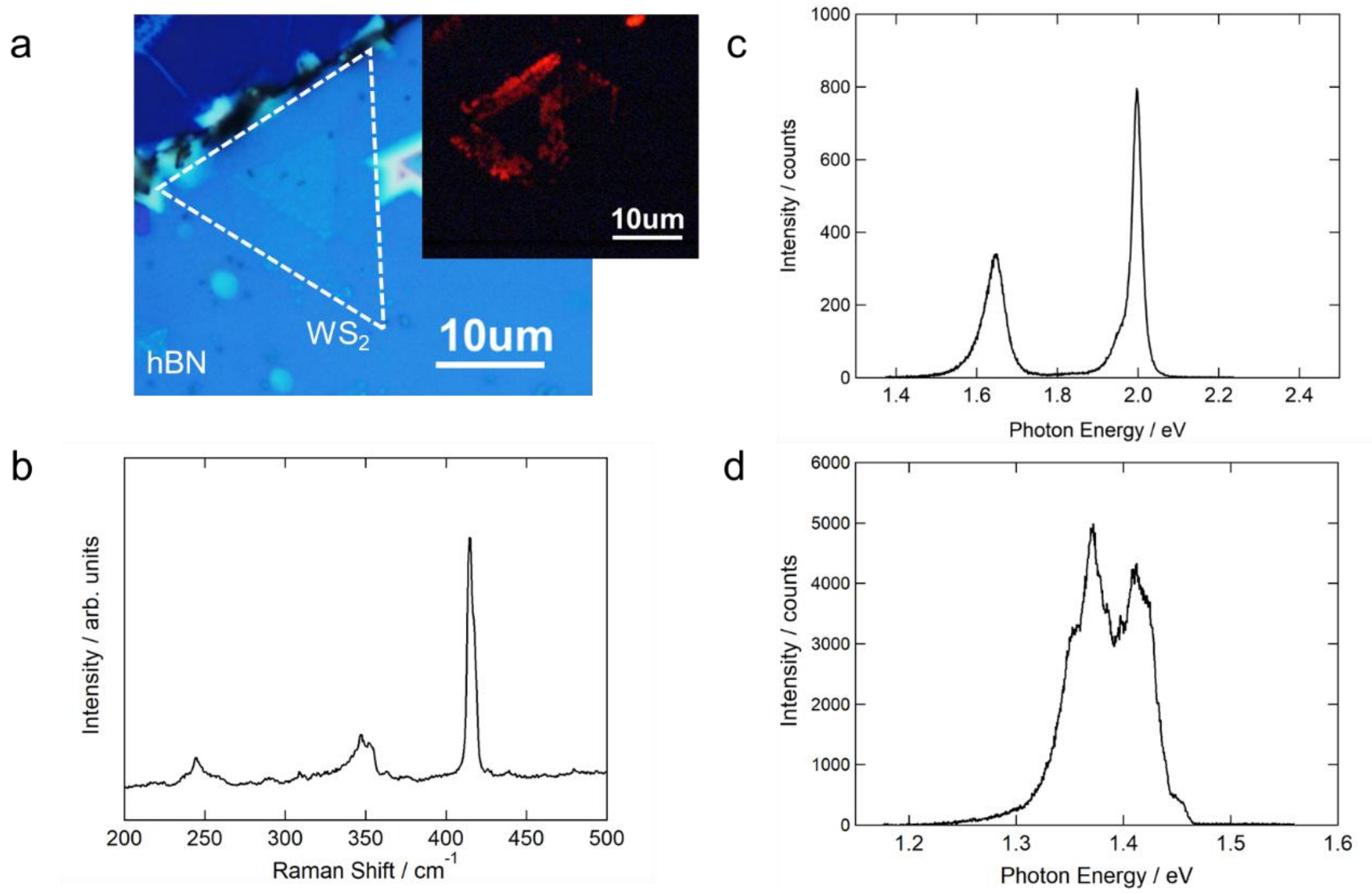

Figure 1. (a) Optical image and PL image of the CVD-grown $WSe_2/WS_2$ sample on hBN. (b) Raman spectrum and (c) PL spectrum of the $WSe_2/WS_2$ sample. These measurements were performed at room temperature. (d) moiré exciton PL spectrum of the $WSe_2/WS_2$ sample measured at 3.5 K.

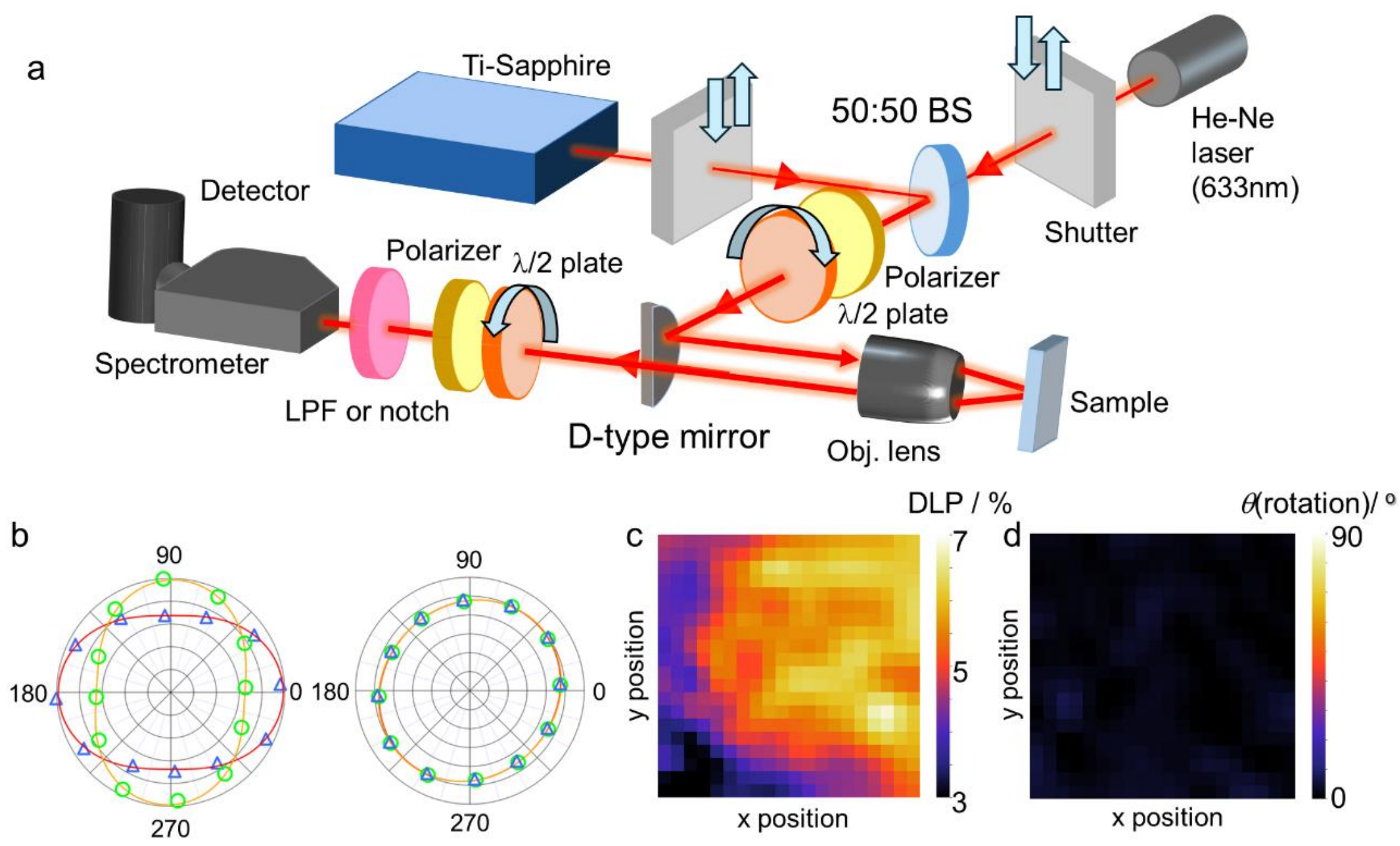

Figure 2. (a) Schematic representation of the optical system used. He-Ne and Ti-sapphire lasers, operating at 633 and 740 nm, respectively, were used for Raman and PL measurements. BS represents a beam splitter. (b) polar plots of PL from a monolayer $WSe_2$ region and a $WSe_2/WS_2$ moiré region, respectively. 90 ° (circle) and 0 ° (triangle) polarizations were used for excitation. (c) and (d) show a 2D map of DLP in % and polarization rotation, $\theta$(rotation), in degrees upon change of excitation polarization. Mapping was performed in a 5.3 × 5.3 μm region of the sample.

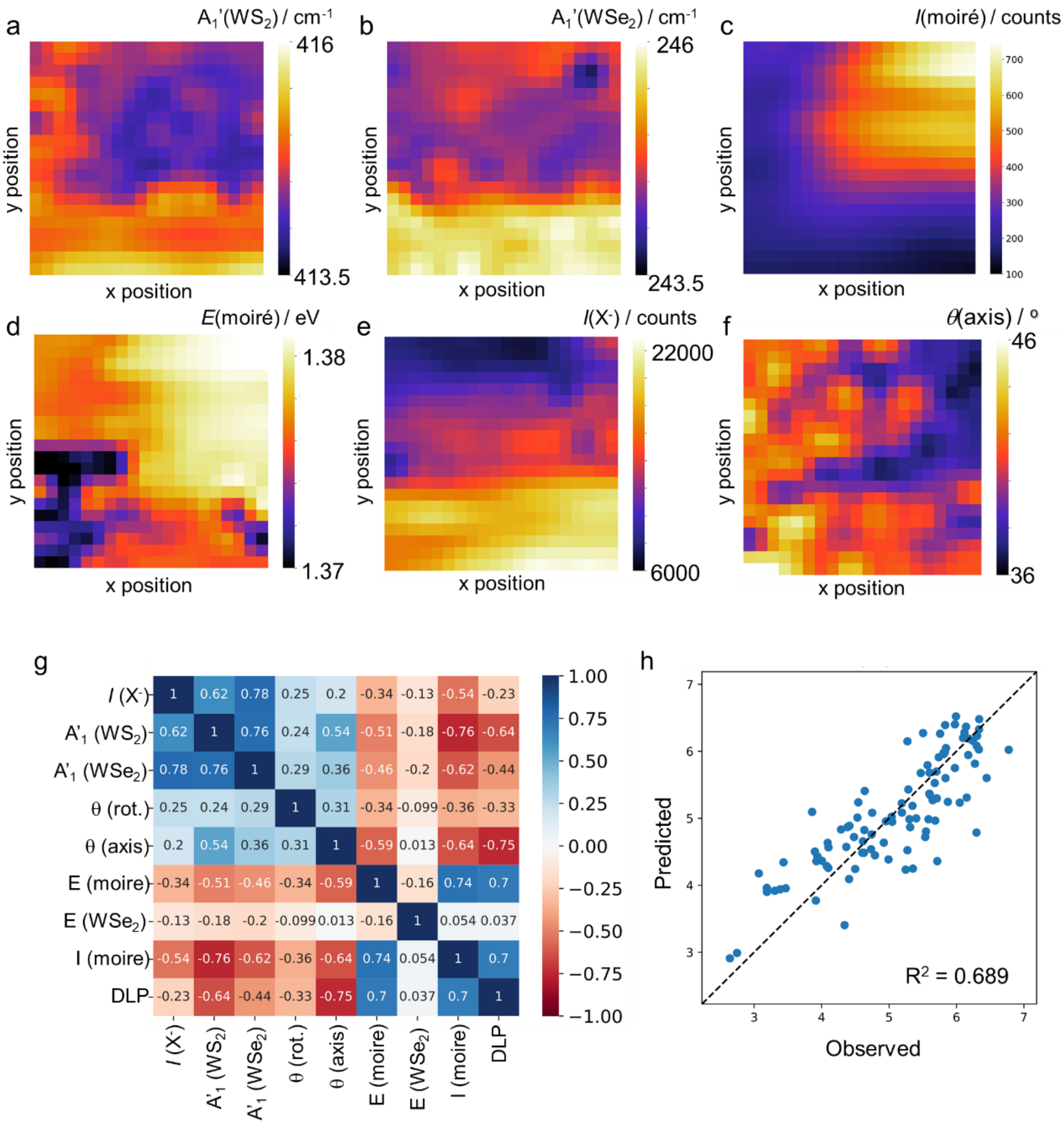

Figure 3. (a)-(f) 2D maps of $WS_2$ A'$_1$ mode, $WSe_2$ A'$_1$ mode, moiré exciton intensity, moiré exciton peak position, trion intensity, and polarization direction, respectively. The units for the color scale for (a)-(f) are $cm^{-1}$, $cm^{-1}$, counts, eV, counts, and degrees, respectively. All maps are taken in the $5.3 \times 5.3$ μm region of the sample used to measure the DLP map shown in Fig. 2(c). All maps are measured at 3.5 K. (g) shows a correlation analysis result, where the numbers indicate Pearson's correlation coefficients for each row-column pair. (h) The relation between predicted and observed DLP. The prediction model was built using linear regression.

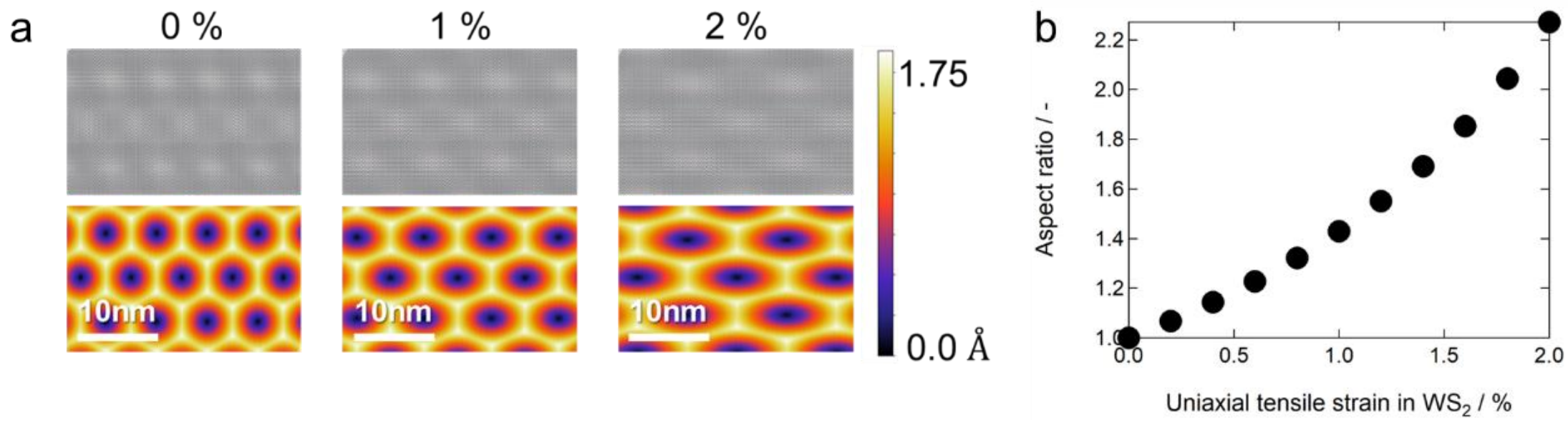

Figure 4. (a) Top figures show structure models of moiré superlattice, $WS_2/WSe_2$, with 0 to 2 % uniaxial tensile strain in $WS_2$. The Poisson ratio of $WS_2$ was used to build the structure models. The bottom figures show the smaller lateral displacement of W atoms between the top and bottom layers. The so-called AA regions, where W atoms sit in almost the same lateral locations, give dark contrast, whereas between the AA regions, the lateral displacement becomes large and gives bright contrasts. (b) The relationship between uniaxial tensile strain in $WS_2$ and the aspect ratio of the moiré site. Aspect ratios were calculated using $d_{\mathrm{WS2}}d_{\mathrm{WSe2}}/(d_{\mathrm{WSe2}}\text{-}d_{\mathrm{WS2}})$ values along horizontal and vertical directions. As the moiré pattern deforms under tensile strain, the aspect ratio increases rapidly.

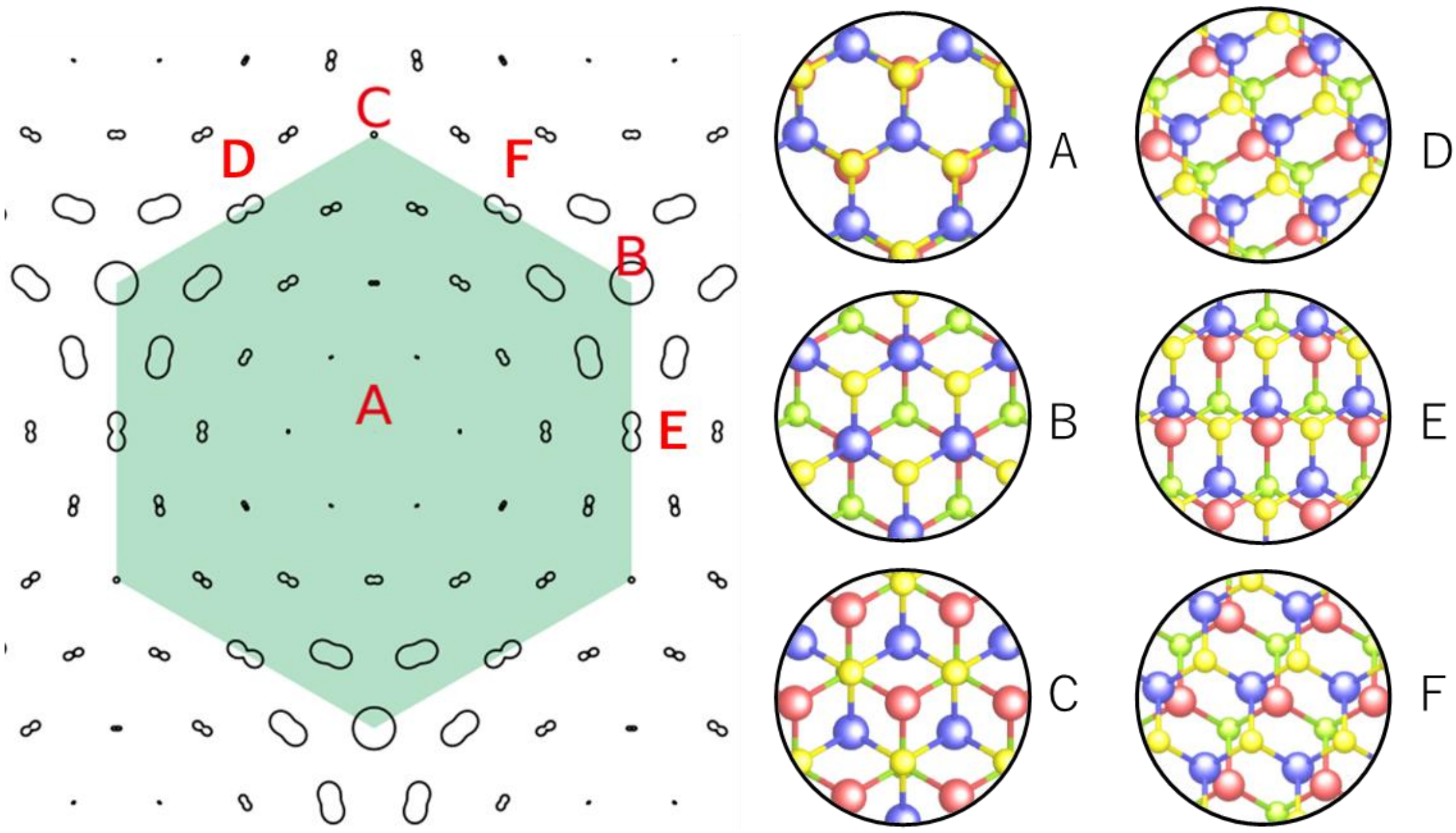

Figure 5. Distribution of the local transition matrix element in the moiré lattice. The green hexagon represents the Wigner–Seitz cell of the moiré lattice, and the local atomic structures around positions A–F are shown on the left. Blue, red, yellow, and green spheres denote W atoms in the top layer, W atoms in the bottom layer, Se atoms, and S atoms, respectively. The circular and peanut-shaped polar plots show the angular dependence of the local transition matrix element. Except at positions A, B, and C, the local transition matrix elements exhibit p-orbital-like angular anisotropy, indicating local linear-polarization selectivity in optical transitions. This local transition-matrix analysis should be distinguished from the circular valley selection rule, which is governed by the global $C_3$ quasi-angular momentum and valley-dependent phase structure of the moiré exciton.

# Supplementary materials

# Strain Correlated Linearly Polarized Photoluminescence in $WS_2/WSe_2$ Moiré Superlattices

*Yuto Urano[1,2], Ryo Tamura[3,4], Yui Tamogami[1,5], Toshikaze Kariyado[1], Yasumitsu Miyata[1], Daichi Kozawa[1], Kenji Watanabe[6], Takashi Taniguchi[1], and Ryo Kitaura[1,2,*]*

*[1] Research Center for Materials Nanoarchitectonics, National Institute for Materials Science, 1-1 Namiki, Tsukuba, Ibaraki 305-0044, Japan*

*[2] Graduate School of Chemical Sciences and Engineering, Hokkaido University, 5, Kita 8 Nishi, Kita-ku, Sapporo, Hokkaido 060-8628, Japan*

*[3] Center for Basic Research on Materials, National Institute for Materials Science, 1-1 Namiki, Tsukuba, Ibaraki 305-0044, Japan*

*[4] Graduate School of Frontier Sciences, The University of Tokyo, 5-1-5 Kashiwanoha, Kashiwa, Chiba 277-8568, Japan*

*[5] Department of Physics, Tokyo Metropolitan University, 1-1 Minamiosawa, Hachioji, Tokyo 192-0397, Japan*

*[6] Research Center for Electronic and Optical Materials, National Institute for Materials Science, 1-1 Namiki, Tsukuba, Ibaraki 305-0044, Japan*

* Correspondence to KITAURA.Ryo@nims.go.jp

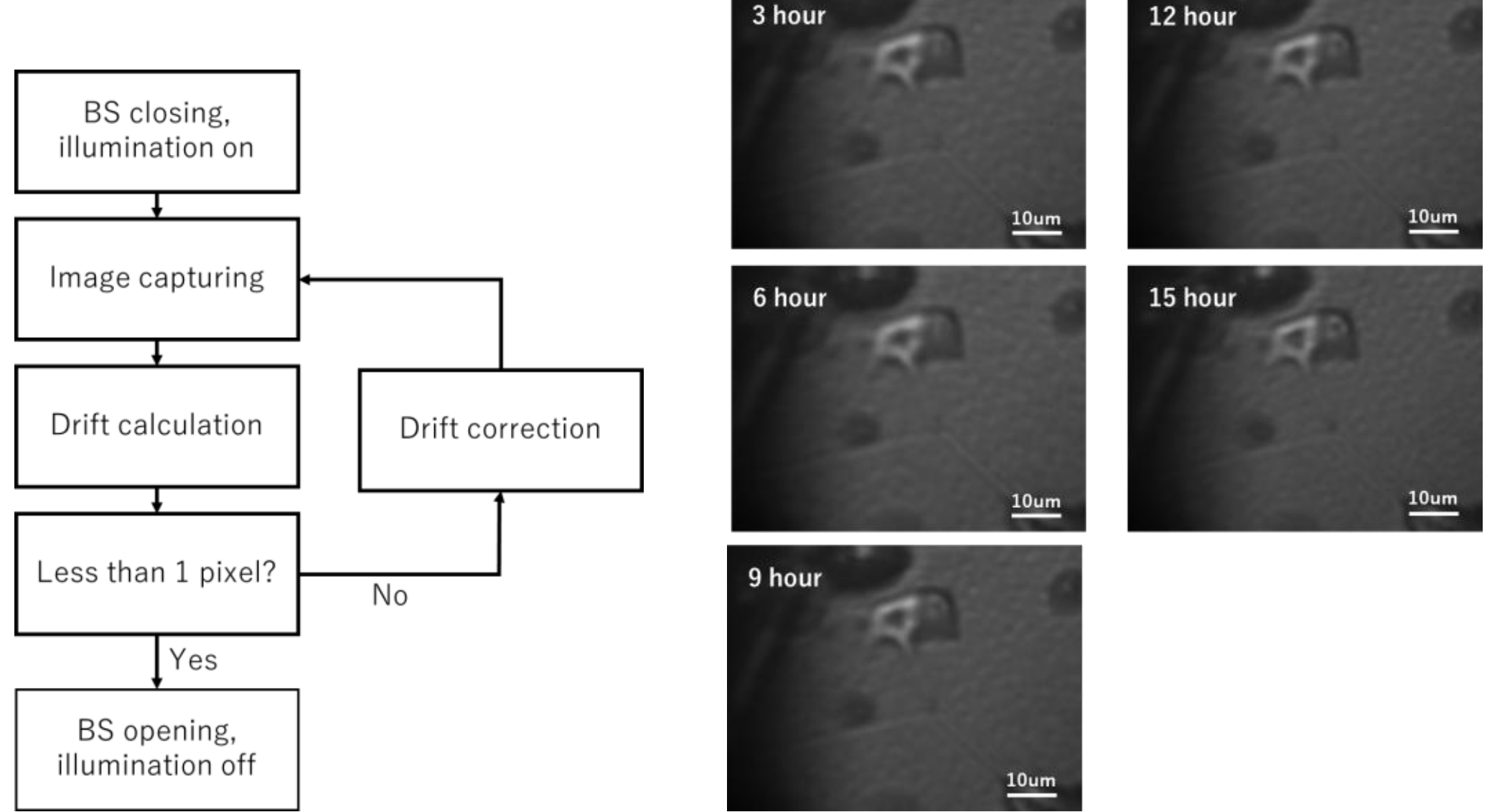

Figure S1 (left) Block diagram illustrating the drift-correction procedure. BS stands for beam shutter. (Right) Series of optical images acquired during a mapping measurement.

Our microspectroscopy system is stable, with a typical spatial drift of less than 1 μm over a three-hour period. However, since mapping measurements can take several days, it is crucial to account for spatial drift to ensure that the spectral maps produced are accurate and meaningful. To correct this drift, we capture an optical image after each line of mapping and calculate the drift using an enhanced correlation method. The calculated drift is measured in pixels and converted into pulses for the stepping motor to perform the correction. All processes—such as closing the beam shutter to cut the excitation laser beam, turning illumination on, capturing an image, calculating the drift, correcting the drift, capturing an additional image to verify if the correction worked, turning illumination off, and reopening the shutter—can be automated using a LabVIEW VI. The correction process is shown in the block diagram in Fig. S1 (left), ensuring drift correction with an accuracy of 1 pixel. Additionally, we incorporate focus correction into the drift correction LabVIEW VI. For focus correction, we direct the excitation laser reflected from the sample through a 50-μm pinhole using a 15-cm lens. The light intensity passing through the pinhole is then measured with a photodiode. Focus correction involves adjusting the sample height to maximize light intensity. This focus correction is performed before drift correction to ensure that the drift correction can be carried out effectively. The figures show optical images of samples collected during a mapping measurement. As seen in the figures, there is almost no shift across all images after the drift correction. This allows us to conduct detailed mapping measurements to analyze the relationship between linear polarization and different quantities.

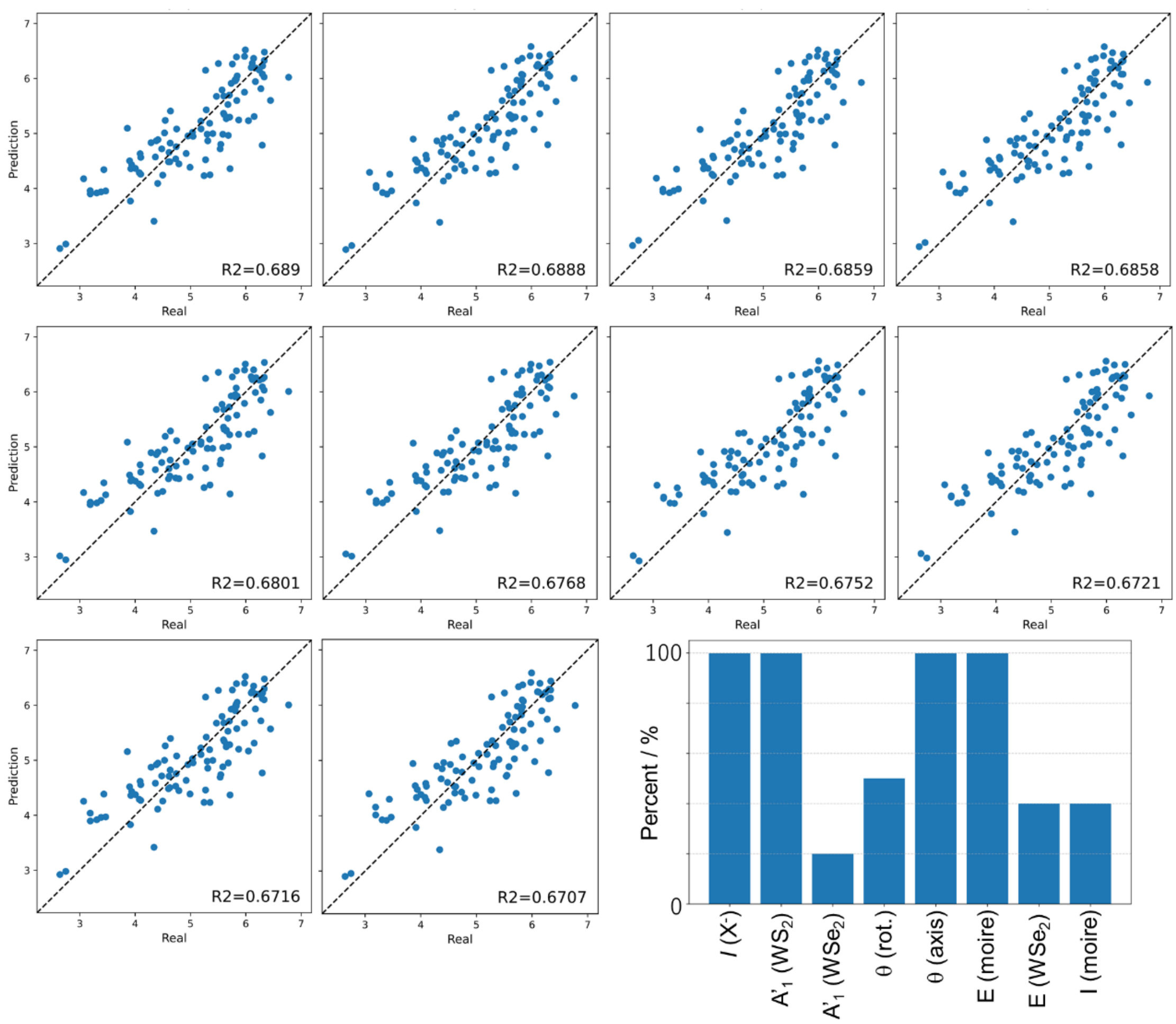

Figure S2 Relations between the observed and predicted DLP. The predictions were made using a top-10 model developed by linear regression. The bar graph shows the physical quantities used in the top-10 models.

We have developed prediction models using all combinations of the eight physical quantities, resulting in a total of 255 (=$2^8$ - 1) models. The plots above illustrate the ten best prediction models among these combinations. The bar graph displays the physical quantities utilized in the top ten models. Notably, $I(X^-)$, $A'_1$ ($WS_2$), $\theta$ (axis), and $E$(moiré) were included in all ten models. This strongly suggests that these four quantities are crucial for explaining the observed DLP in the moiré sample.

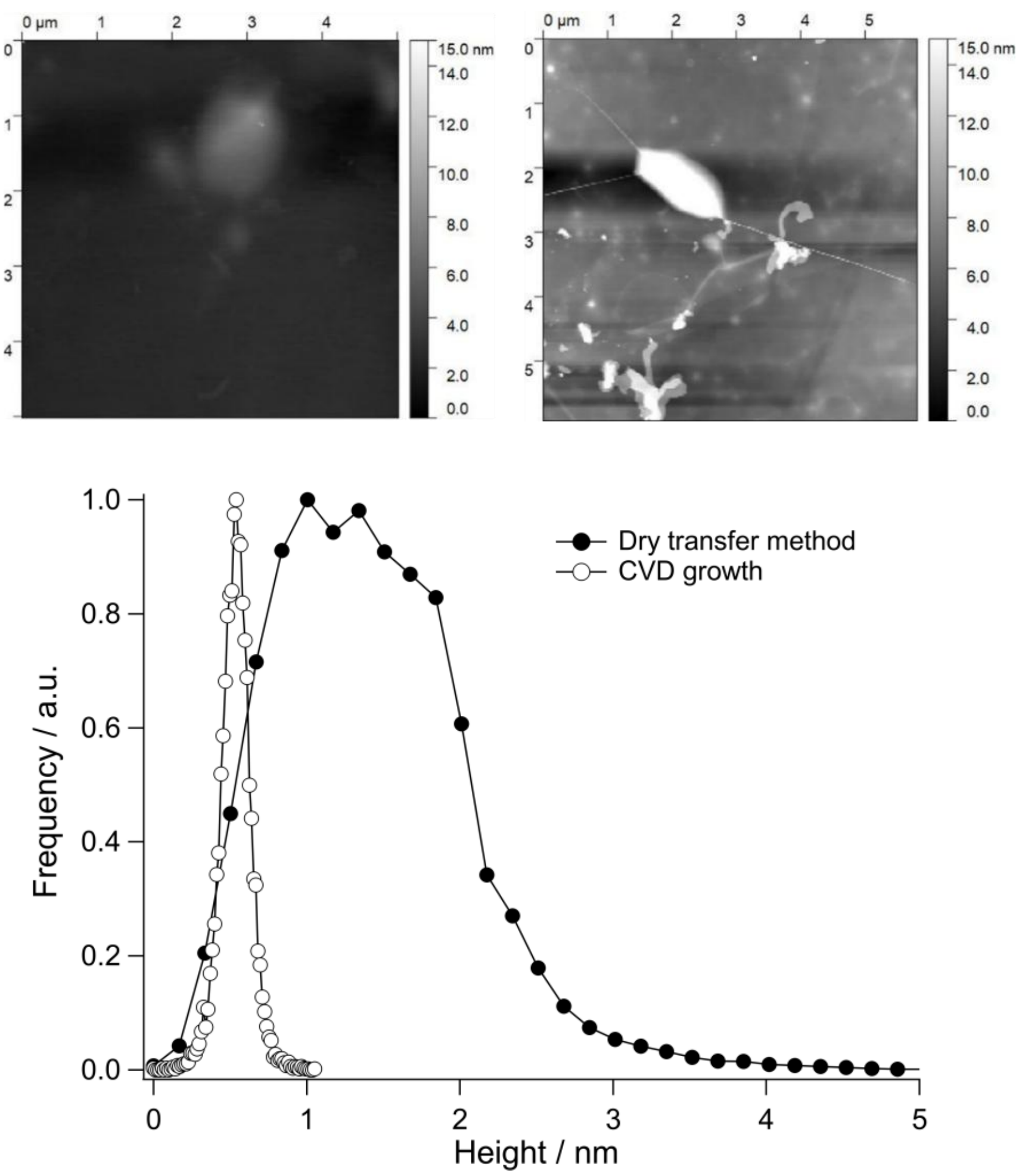

Figure S3. AFM images of (top left) the CVD-grown $WSe_2/WS_2$ sample and (top right) a dry-transferred sample. (Bottom) Height distributions of the CVD-grown and dry-transferred samples. These moiré samples are encapsulated between hBN flakes.

These AFM images are representative of each sample type, and the high regions correspond to bubbles trapped between layers. In the dry-transferred sample, bubbles can be introduced during each transfer step used to assemble the $hBN/WSe_2/WS_2/hBN$ stack. In the CVD-grown sample, bubbles can form when a top hBN flake is placed on the $WS_2/WSe_2/hBN$ surface to protect the heterostructure. The height distributions clearly show that, apart from the encapsulated bubbles, the surface of the CVD-grown sample remains relatively flat, consistent with the formation of a clean interface between $WS_2$ and $WSe_2$. In contrast, the dry-transferred sample exhibits a much broader height distribution, attributed to bubbles trapped between layers during transfer.

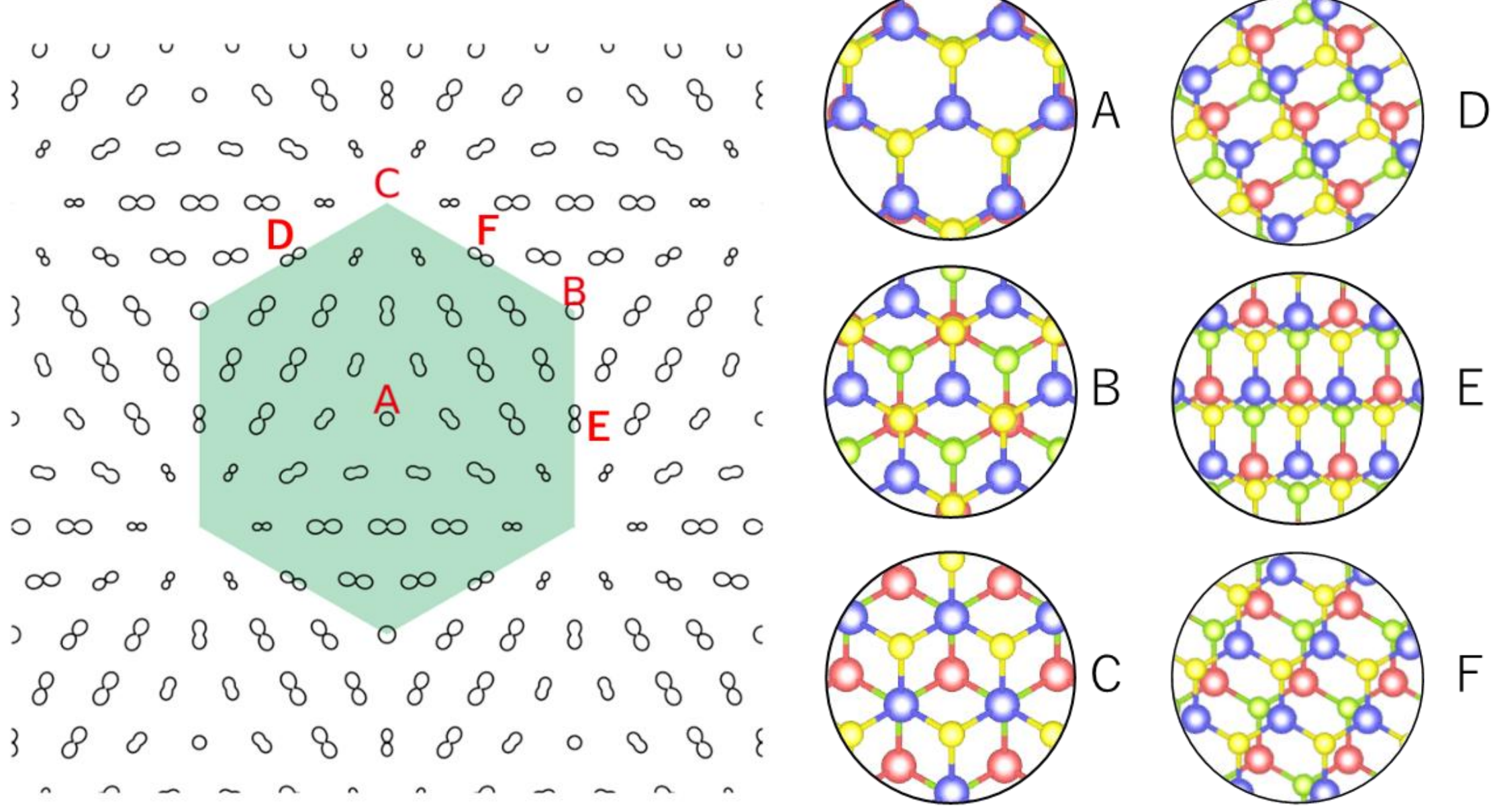

Figure S4. Distribution of the local transition matrix element in the moiré lattice.

The green hexagon represents the Wigner-Seitz cell of the moiré lattice, whose local structures around A-F are shown on the left. Blue, red, yellow, and green spheres denote W (top layer), W (bottom layer), Se, and S atoms, respectively. Circles and peanut-shaped figures are polar plots of the local transition matrix element. Except for the locations A, B, and C, all transition matrix elements display a p-orbital-like shape, indicating possible local polarization selectivity in optical transitions.

**Supporting Note S5. Local transition-matrix analysis and circular valley selection rule**

The local transition-matrix analysis shown in Fig. 5 should be distinguished from the circular valley selection rule of moiré excitons. Figure 5 evaluates the angular dependence of the local transition matrix element at different atomic registries within the moiré unit cell. This analysis is useful for visualizing local linearly polarized emission components and for discussing how these local components can cancel under $C_3$ symmetry or remain finite when strain makes the $C_3$-related emission weights inequivalent.

In contrast, the circular valley readout is governed by the global $C_3$ quasi-angular momentum of the moiré exciton. In a $C_3$-symmetric lattice, the K point remains invariant under $C_3$ rotation up to a reciprocal lattice vector. Consequently, valley Bloch states at K can be classified according to their $C_3$ eigenvalues. The circular optical selection rule arises from the conservation of $C_3$ quasi-angular momentum modulo 3, incorporating the $C_3$ transformation properties of both the electron and hole valley Bloch functions, as well as the phase structure of the exciton envelope across the moiré unit cell. Such a global symmetry property is not captured by a purely local transition-matrix map. Figure 5, therefore, does not aim to assess whether the circularly polarized valley readout is preserved or suppressed.

The role of Fig. 5 in the present work is more specific: it illustrates that local optical transitions can contain linearly polarized components. In a $C_3$-symmetric moiré lattice, three $C_3$-related local emission channels contribute with equal weights, and their linearly polarized components cancel in the far field. When residual strain distorts the moiré atomic registry and the exciton envelope, these local weights become inequivalent. This incomplete cancellation provides a symmetry-based route to finite far-field DLP, as discussed in Supporting Note S6.

**Supporting Note S6. Minimal $C_3$-breaking model for finite far-field DLP**

Here, we introduce a minimal symmetry-based model to show how a finite far-field DLP can emerge from weak $C_3$ symmetry breaking in a moiré exciton system. This model is not intended to provide a full microscopic calculation of the exciton wavefunction or optical matrix elements. Rather, it clarifies the symmetry requirement for the incomplete cancellation of local linearly polarized emission components. We consider three local emission channels related by $C_3$ rotation within a moiré unit cell. Each channel is assumed to have a local linearly polarized emission component, as shown by the DFT calculation. Using the complex electric-field amplitudes along the x and y axes, $E_x$ and $E_y$, the linear Stokes parameters are written as

$$Q = S_1 = \langle |E_x|^2 - |E_y|^2 \rangle,$$

$$U = S_2 = 2\,\mathrm{Re}\langle E_x E_y^{*} \rangle,$$

where <...> denotes time or ensemble averaging. Thus, $Q$ represents the intensity imbalance between x- and y-polarized components, whereas $U$ represents the intensity imbalance between +45° and −45° linear polarization components. The complex linear-polarization parameter used in the minimal model is therefore defined as

$$\mathcal{P} = Q + \mathrm{i}U = S_1 + \mathrm{i}S_2.$$

If the local polarization axis of the j-th channel is $\varphi_j$, its contribution is written as

$$\mathcal{P}_j = p_0\, w_j \exp(\mathrm{i}2\varphi_j)$$

Here, $p_0$ is the local degree of linear polarization, and $w_j$ is the effective emission weight of the j-th channel. The factor of 2 in $\exp(\mathrm{i}2\varphi_j)$ reflects the fact that a linear polarization axis is unchanged under $\varphi_j \rightarrow \varphi_j + \pi$. In the far-field DLP measurements, the sum of each contribution is observed.

$$\mathcal{P} = p_0\, \Sigma_j\, w_j \exp(\mathrm{i}2\varphi_j)$$

Below, we calculate $\mathcal{P}$ for both the $C_3$-symmetric case and the case with weakly broken $C_3$ symmetry.

**1. The $C_3$-symmetric case**

The three local polarization axes are related by 120° rotations,

$$\varphi_j = \varphi_0 + 2\pi(j-1)/3,\ j = 1, 2, 3,$$

and the three channels have identical weights,

$$w_1 = w_2 = w_3 = w_0$$

The total far-field linear polarization is then

$$\mathcal{P} = p_0\, w_0\, \Sigma_j \exp(\mathrm{i}2\varphi_j) = 0.$$

Thus, even if each local channel has a finite linearly polarized component, the far-field DLP vanishes in the $C_3$-symmetric limit because the three contributions cancel exactly.

**2. The case with weakly broken $C_3$ symmetry**

We introduce weak $C_3$ symmetry breaking through an effective parameter $\varepsilon$. Physically, $\varepsilon$ represents the strain-induced inequivalence of the three local emission channels, causing inequivalence in $w_j$. This inequivalence can arise from distortion of the moiré pattern, modification of the local optical matrix elements, and/or distortion of the exciton envelope function. Specifically, $w_j$ can be written as follows.

$$w_j(\varepsilon) = \int_{\Omega_j(\varepsilon)} d^2r \, |\chi_\varepsilon(r)|^2 \, g_j(r, \varepsilon),$$

where $\Omega_j(\varepsilon)$ is the emission-contributing region associated with the j-th local channel, $\chi_\varepsilon(r)$ is the strain-distorted exciton envelope function, and $g_j(r, \varepsilon)$ represents the local oscillator strength. In the absence of strain, $C_3$ symmetry makes the three weights equivalent. When strain distorts the moiré pattern, the regions $\Omega_j(\varepsilon)$, the envelope distribution $|\chi_\varepsilon(r)|^2$, and the local oscillator strength can become inequivalent among the three channels. All of these effects are phenomenologically captured by the effective $C_3$-breaking parameter $\varepsilon$.

We assume that the effective emission weight of each local channel responds linearly to the anisotropic component of the strain tensor. Let $\mathbf{e}_j = (\cos\varphi_j, \sin\varphi_j)$ be the unit vector along the local emission axis of the j-th channel. We write

$$w_j = w_0 \, [1 + \lambda \, \mathbf{e}_j^T \mathbf{u}_{dev} \boldsymbol{e}_j],$$

where $\mathbf{u}_{dev}$ is the deviatoric component of the in-plane strain tensor, and $\lambda$ is a phenomenological coupling constant. The hydrostatic component of strain changes all three weights equally and therefore does not generate a net far-field DLP to first order.

For a uniaxial distortion with principal axis $\psi_s$, the deviatoric strain tensor can be written as

$$\mathbf{u}_{dev} = u_s \begin{pmatrix} \cos 2\psi_s & \sin 2\psi_s \\ \sin 2\psi_s & -\cos 2\psi_s \end{pmatrix}$$

where $u_s = (u_{parallel} - u_{perpendicular})/2$ is the magnitude of the anisotropic strain component. This gives

$$\mathbf{e}_j^T \mathbf{u}_{dev} \boldsymbol{e}_j = u_s \cos 2(\varphi_j - \psi_s).$$

Therefore,

$$w_j = w_0 \, [1 + \lambda u_s \cos 2(\varphi_j - \psi_s)].$$

This form expresses the physical assumption that local emission channels aligned parallel and perpendicular to the strain-induced distortion acquire different effective emission weights. Equivalently, by defining $\varepsilon = \lambda u_s/2$, we obtain

$$w_j = w_0 \, [1 + 2\varepsilon \cos 2(\varphi_j - \psi_s)].$$

The total polarization thus becomes

$$\mathcal{P} = p_0 \, \Sigma_j \, w_j \exp(i2\varphi_j).$$

The $C_3$-symmetric part again cancels. Keeping only the term linear in $\varepsilon$ gives

$$\mathcal{P} = 3\, p_0\, w_0\, \varepsilon \exp(\mathrm{i}2\psi_s).$$

The total emission intensity is

$$S_0 = \Sigma_j\, w_j = 3w_0 + O(\varepsilon^2).$$

Since $S_0 = 3w_0 + O(\varepsilon^2)$, the denominator does not modify the leading linear dependence, and we obtain the far-field DLP as

$$\mathrm{DLP} = |\mathcal{P}|/S_0 = p_0|\varepsilon| + O(\varepsilon^2).$$

This result shows that finite DLP appears linearly with the effective $C_3$-breaking parameter. Therefore, a weak perturbation that makes the three $C_3$-related local emission weights inequivalent is sufficient to generate a finite far-field DLP.

The effective channel weights $w_j$ can be interpreted as a coarse-grained representation of the spatially integrated emission contribution. This point can be seen from a continuous expression for the complex linear-polarization component,

$$\mathcal{P}(\varepsilon) = \int_{\mathrm{cell}} \mathrm{d}^2r\, |\chi_\varepsilon(r)|^2\, m_\varepsilon(r),$$

where $\chi_\varepsilon(r)$ is the strain-dependent exciton envelope function and $m_\varepsilon(r)$ is the local complex linear-polarization density, including the local optical matrix element and polarization axis. For weak $C_3$ breaking,

$$|\chi_\varepsilon(r)|^2 = |\chi_0(r)|^2 + \varepsilon\, \delta\rho_\chi(r),$$

$$m_\varepsilon(r) = m_0(r) + \varepsilon\, \delta m(r).$$

We define $\rho_\chi(r) = |\chi_\varepsilon(r)|^2$ as the exciton-envelope probability density, and $\delta\rho_\chi(r)$ represents the first-order change in the spatial weighting of the exciton envelope.

Substitution gives

$$\mathcal{P}(\varepsilon) = \int_{\mathrm{cell}} \mathrm{d}^2r\, |\chi_0(r)|^2\, m_0(r) + \varepsilon \int_{\mathrm{cell}} \mathrm{d}^2r\big(\, \delta\rho\chi(r) m_0(r) + |\chi_0(r)|^2 \delta \mathrm{m}(\mathrm{r})\, \big) + O(\varepsilon^2).$$

The first term vanishes in the $C_3$-symmetric limit. The two linear terms show that finite far-field DLP can arise either from strain-induced distortion of the exciton envelope, which changes the spatial weighting of local emission components, or from strain-induced modification of the local optical matrix element. Thus, the phenomenological weight modulation used in the three-channel model captures both effects at the lowest order in the effective $C_3$-breaking parameter.

This minimal model shows that strain does not need to create a completely new optical selection rule. It only makes the three $C_3$-related local emission channels inequivalent. Once this happens, the local linearly polarized components no longer cancel completely, and a finite far-field DLP appears. Since small lattice strain can be amplified into a larger distortion of the moiré pattern, even weak residual strain can generate a measurable DLP in moiré exciton emission.

**Supporting Note S7. Weak $C_3$-breaking and VDF**

Weak $C_3$ symmetry breaking can affect linear- and circular-polarization readout in qualitatively different ways. In the valley basis, circularly polarized emission is associated mainly with valley population imbalance, whereas linearly polarized emission is associated with the coherent superposition of the K and −K valleys. Therefore, an additional strain-induced linear-polarization channel can appear without necessarily eliminating the circular contrast used to read out valley polarization. Phenomenologically, the optical dipole of a K-valley exciton can be written as

$$\boldsymbol{d}_{\mathrm{K}} = d_0(\mathbf{e}_+ + \eta \mathbf{e}_-),$$

where $\mathbf{e}_+$ and $\mathbf{e}_-$ denote right- and left-circular polarization components, and $\eta$ is a small helicity-mixing parameter induced by symmetry breaking. The phenomenological dipole expression $\boldsymbol{d}_{\mathrm{K}}$ is directly related to the perturbative optical matrix elements. We define the transition dipole vector of a K-valley exciton as

$$\boldsymbol{d}_{\mathrm{K}} = \langle G|\boldsymbol{d}|X_{\mathrm{K}}\rangle,$$

where $\boldsymbol{d}$ is the electric dipole operator. Its projections onto the circular polarization basis give the optical matrix elements,

$$M_\sigma = \langle G|\mathbf{e}_\sigma^* \cdot \boldsymbol{d}|X_{\mathrm{K}}\rangle = \mathbf{e}_\sigma^* \cdot \boldsymbol{d}_{\mathrm{K}}.$$

Thus, $\boldsymbol{d}_{\mathrm{K}}$ can be expanded as

$$\boldsymbol{d}_{\mathrm{K}} = M_+\mathbf{e}_+ + M_-\mathbf{e}_-.$$

In the ideal $C_3$-symmetric limit, the K-valley exciton couples only to one circular helicity, so that

$$M_+^{(0)} = d_0, \ \mathrm{M}_-^{(0)} = 0.$$

When weak $C_3$-symmetry breaking is introduced through $H = H_0 + \lambda V$, where $\lambda V$ represents a weak $C_3$-breaking perturbation. We denote the total first-order correction to the optical matrix element by $\delta M_\sigma$, including corrections from the perturbed exciton state, the ground state, and, if necessary, the optical transition operator. The allowed matrix element thus becomes

$$M_+ = d_0 + \lambda \delta M_+ + O(\lambda^2),$$

whereas the previously forbidden opposite-helicity matrix element becomes

$$M_- = \lambda \delta M_- + O(\lambda^2).$$

Therefore,

$$\boldsymbol{d}_{\mathrm{K}} = (d_0 + \lambda \delta M_+)\mathbf{e}_+ + \lambda \delta M_- \mathbf{e}_- + O(\lambda^2).$$

By absorbing the small correction to the allowed component into the definition of $d_0$ and defining

$$\eta = M_-/M_+ = \lambda \delta M_-/d_0 + O(\lambda^2),$$

we obtain the compact phenomenological form

$$\boldsymbol{d}_{\mathrm{K}} = d_0(\mathbf{e}_+ + \eta \mathbf{e}_-).$$

Thus, η is the ratio of the weakly allowed opposite-helicity amplitude to the dominant allowed helicity amplitude. Fermi's golden rule then gives $I_+ \propto |d_0|^2$ and $I_- \propto |d_0|^2|\eta|^2$. The leakage into the opposite circular polarization, therefore, appears only at order $|\eta|^2$, even though the forbidden optical matrix element itself is generated at first order in the symmetry-breaking perturbation. Specifically, the degree of circular polarization (DCP) can be written as

$$\mathrm{DCP} = (I_+ - I_-)/(I_+ + I_-) \simeq 1 - 2|\eta|^2$$

In contrast, a strain-induced imbalance among $C_3$-related local emission weights can generate a finite linear-polarization component to first order in the effective $C_3$-breaking parameter, as discussed in Supporting Note S6. This simple consideration shows that weak residual symmetry breaking can produce a measurable DLP while circular valley readout remains approximately intact. The key implication is therefore not that valley-selective optical selection rules fail in moiré heterobilayers, but that finite DLP alone is not sufficient evidence for valley coherence.